\documentclass[aps,showpacs,floatfix,twocolumn]{revtex4}
\usepackage[brazil,english]{babel}
\usepackage[T1]{fontenc}
\usepackage{amsmath,amsfonts,amssymb,float,graphicx,epsfig,color,times,bbm}

\usepackage{hyperref}
\begin{document}
\title{Active control of qubit-qubit entanglement evolution}
\author{J. F. Leandro}
\affiliation{Departamento de Física, Universidade Estadual de Ponta Grossa - Campus Uvaranas, 84030-900 Ponta Grossa, Paraná, Brazil}
\author{A. S. M. de Castro}
\affiliation{Departamento de Física, Universidade Estadual de Ponta Grossa - Campus Uvaranas, 84030-900 Ponta Grossa, Paraná, Brazil}
\author{P. P. Munhoz}
\affiliation{Departamento de Física, Universidade Estadual de Ponta Grossa - Campus Uvaranas, 84030-900 Ponta Grossa, Paraná, Brazil}
\author{F. L. Semi\~ao}
\affiliation{Departamento de Física, Universidade Estadual de Ponta Grossa - Campus Uvaranas, 84030-900 Ponta Grossa, Paraná, Brazil}
\begin{abstract}
In this work, we propose a scheme to design the time evolution of the entropy of entanglement between two qubits. It is shown an explicit accurate solution for the inverse problem of determining the time dependence of the coupling constant from a user-defined dynamical entanglement function. Such an active control of entanglement can be implemented in many different physical implementations of coupled qubits, and we briefly comment on the use of interacting flux qubits.
\end{abstract}
\maketitle
\section{Introduction}
In quantum information science, entanglement is commonly recognized as a resource that may be used to perform novel information processing tasks \cite{ms}. A famous example of such a task is quantum teleportation \cite{tele}. Two parties sharing a pure maximally entangled pair of qubits can perform the perfect transmission of the state of a third qubit using just local operations and classical communication. This would be impossible with the use of classical communication alone. This view of entanglement as a resource has triggered the investigation of measures of entanglement. Complete and didactic reviews of the main entanglement measures and their fundamental properties can be found in \cite{ms,h}.

The common approach to generate entanglement between qubits or more complicated quantum systems consists of letting the system evolve under some time-independent interaction Hamiltonian. In general, the total time of evolution and the initial preparation of the state of the system are the ingredients varied to achieve the desired final entangled state. In this passive approach, the system entanglement follows a unique time evolution once the initial state is fixed. On the other hand, the use of time-dependent Hamiltonians could allow for a higher level of control over the time evolution of entanglement. In this case, even if the initial state is fixed, the variation of the coupling between the subsystems during time evolution may lead to many different paths for the entanglement to follow during the same interval of time.

In this article, we propose the use of time-dependent Hamiltonians to control the entanglement creation between qubits. We do it in an active way, i.e. after defining the \emph{shape} wanted for the entanglement time evolution, the coupling magnitude is dynamically varied in order to achieve the expected behavior. The use of time-dependent Hamiltonians to establish entanglement is in fact gaining  impetus in the last years. In \cite{er}, time-dependent spin-spin couplings are used for creation of
long-distance entanglement where it is predicted an increase of entanglement between the first and last spin of a chain whenever the ac part of the coupling has a frequency matching the Zeeman splitting. In \cite{jcmd}, sine and rectangular field frequency modulation is shown to be favorable to improve, enhance and stabilize the degree of the atom-field entanglement in a typical atom-cavity model. In \cite{act}, optimal control theory is used to determine the coupling modulation that leads to
maximum logarithmic negativity for a pair of opposite oscillators in a harmonic chain. We now provide a solution for the \emph{entanglement design} between qubits through the choice of the appropriate wave form of the coupling intensity. This paper is organized as follows. In Section \ref{ansatz}, we provide an approximate solution for the inverse problem of obtaining the time dependent coupling from a desired time dependent entanglement function. In Section \ref{decoherence}, we discuss the meaning of these different paths and find applications when the qubits are subjected to decoherence. We summarize and conclude in Section \ref{conclusions}.
\section{Statement and Solution of the Problem}\label{ansatz}
\subsection{Ansatz}
In order to illustrate the general idea behind \emph{entanglement design}, let us consider the simple two-qubit XY Hamiltonian
\begin{eqnarray}\label{H}
H(t)=\frac{\lambda(t)}{2}(\sigma_1^x\sigma_2^x+\sigma_1^y\sigma_2^y),
\end{eqnarray}
where $\lambda(t)$ is a time-dependent coupling magnitude between resonant qubits and $\sigma$'s are usual Pauli matrices. In the computational basis, if the initial state of the system is $|\psi(0)\rangle=|01\rangle$, the evolved state will be
\begin{eqnarray}
|\psi(t)\rangle=\cos\eta(t)|01\rangle-i\sin\eta(t)|10\rangle,\label{ev}
\end{eqnarray}
where
\begin{eqnarray}
\eta(t)=\int_0^t\lambda(\tau)d\tau {\bf \label{eu}}
\end{eqnarray}
is the area of the interaction pulse. Since the evolved state (\ref{ev}) is pure, the entanglement between the qubits can be quantified by means of the entropy of entanglement \cite{ee}. The result is the composite function
\begin{eqnarray}
S[\eta(t)]=-c[\eta(t)]\log_2\{c[\eta(t)]\}-s[\eta(t)]\log_2\{s[\eta(t)]\}, \label{ee}
\end{eqnarray}
where $c[\eta(t)]=\cos^2\eta(t)$ and $s[\eta(t)]=\sin^2\eta(t)$. Our goal is to \textit{draw} the time evolution of the entanglement function. Mathematically, we would like to obtain the area pulse $\eta(t)$ that satisfies the equation $S[\eta(t)]=f(t)$ for an arbitrary function $f(t)$. This function $f(t)$ is the shape we want the entanglement to have. For two qubits, this function is limited to assume values between $0$ and $1$, and for the initial preparation considered here, $f(t)$ must also obey $f(0)=0$. Should such $\eta(t)$ be found, an experimentalist able to control the interaction area pulse will not only gain control over the amount of entanglement created from $t=0$ to $t=T$, but also on how the entanglement actually evolved during this time interval.

We provide now an approximate analytical solution for this problem, and illustrate its validity with some particular choices of $f(t)$. Clearly, the pulse area will have to be a function of $f(t)$, otherwise the problem does not make sense. Consequently, the entanglement itself might be considered a function of $f(t)$. When thinking this way, the solution of $S[f(t)]=f(t)$ is nothing more than an attempt to linearize the entropy of entanglement, regarding $f(t)$ as the independent variable. This linearization (or approximate linearization) is precisely our strategy to solve the problem.

We start by considering the linear entropy $S_L=2(1-{\rm{Tr}\rho^2_R})$ where $\rho_R$ is the reduced state obtained by tracing out one of qubits. Using (\ref{ev}), we obtain $S_L=\sin^2[2 \eta(t)]$. The design of the linear entropy can be performed exactly because the solution of $S_L=f(t)$ is just
\begin{eqnarray}
\eta(t)=\frac{1}{2}\arcsin[f(t)^{1/2}].\label{sle}
\end{eqnarray}
However, our main goal is to draw the entropy of entanglement. We then slightly modify the solution (\ref{sle}) trying to compensate the information lost when moving from the entropy of entanglement to the linear entropy. We propose the following \emph{ansatz}
\begin{eqnarray}
\eta(t)=\frac{1}{2}\arcsin[f(t)^{q/2}],\label{an}
\end{eqnarray}
where $q$ is an adjustable parameter. We are now going to optimize the value of $q$ such that $S[\eta(t)]$ becomes as close as possible to $f(t)$. We have already mentioned that this is equivalent to approximate $y_1(f;q)=S(f)$ to $y_2(f)=f$ when we think of $f$ as an independent variable $(f\in[0,1])$. Let $M=C[0,1]$ the set of all continuous functions on $[0,1]$. The functions $y_1(f;q)$ and $y_2(f)$ are elements of $M$. We can define a metric (distance) for this set which for $y_1$ and $y_2$ reads \cite{rs}
\begin{eqnarray}
d(q)=\int_0^1|y_1(f;q)-y_2(f)|df.
\end{eqnarray}
\begin{figure}[h]
 \centering\includegraphics[width=0.8\columnwidth]{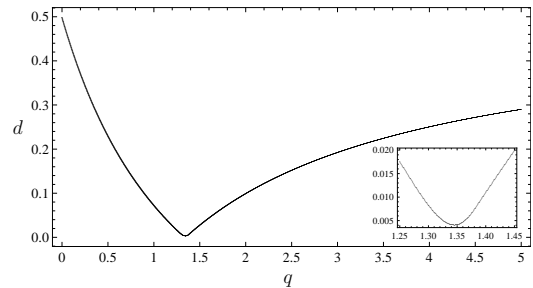}
 \caption{Distance between the entropy of entanglement and a perfect straight line as a function of the adjustable parameter $q$. The inset shows the region where lies the minimum of $d(q)$.}
 \label{figq}
\end{figure}
In Fig.\ref{figq}, this distance is plotted as a function of the adjustable parameter $q$. It is clear that there is a value of $q$ ($q=1.345$) that minimizes the distance between both curves to less than $5\times 10^{-3}$. In order to analyze to which extent this is a good approximation, we show in Fig.\ref{fig1} the entropy of entanglement (\ref{ee}) as a function of $f(t)$ using our ansatz (\ref{an}). An exact solution would lead to a perfect straight line with slope equal to $1$ and crossing the origin (dashed line). One can see that our ansatz, with that value of $q$, is indeed a very good solution of the problem. This plot shows that for small values of $f(t)$, i.e. small entanglement, the solution works less well as indicated by the slight deviation from a perfect straight line. However, as we are going to see from explicit examples, this deviation does not do much harm to the design of entanglement. The time dependence of the coupling is found from the first derivative of (\ref{an}) with respect to time
\begin{eqnarray}
\lambda(t)=\frac{d\eta(t)}{dt}= \frac{1}{4}\frac{qf(t)^{-1+q/2}}{\sqrt{1-f(t)^q}}\frac{df(t)}{dt}.\label{c}
\end{eqnarray}
\begin{figure}[h]
 \centering\includegraphics[width=0.8\columnwidth]{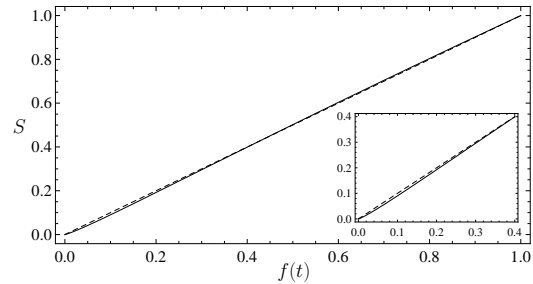}
 \caption{The designed entropy of entanglement $S$ as a composed function of the target function $f(t)$. The inset shows the region where the ansatz deviates more from a perfect straight line.}
 \label{fig1}
\end{figure}

\begin{figure}[h]
\begin{center}
\includegraphics[width=0.8\columnwidth]{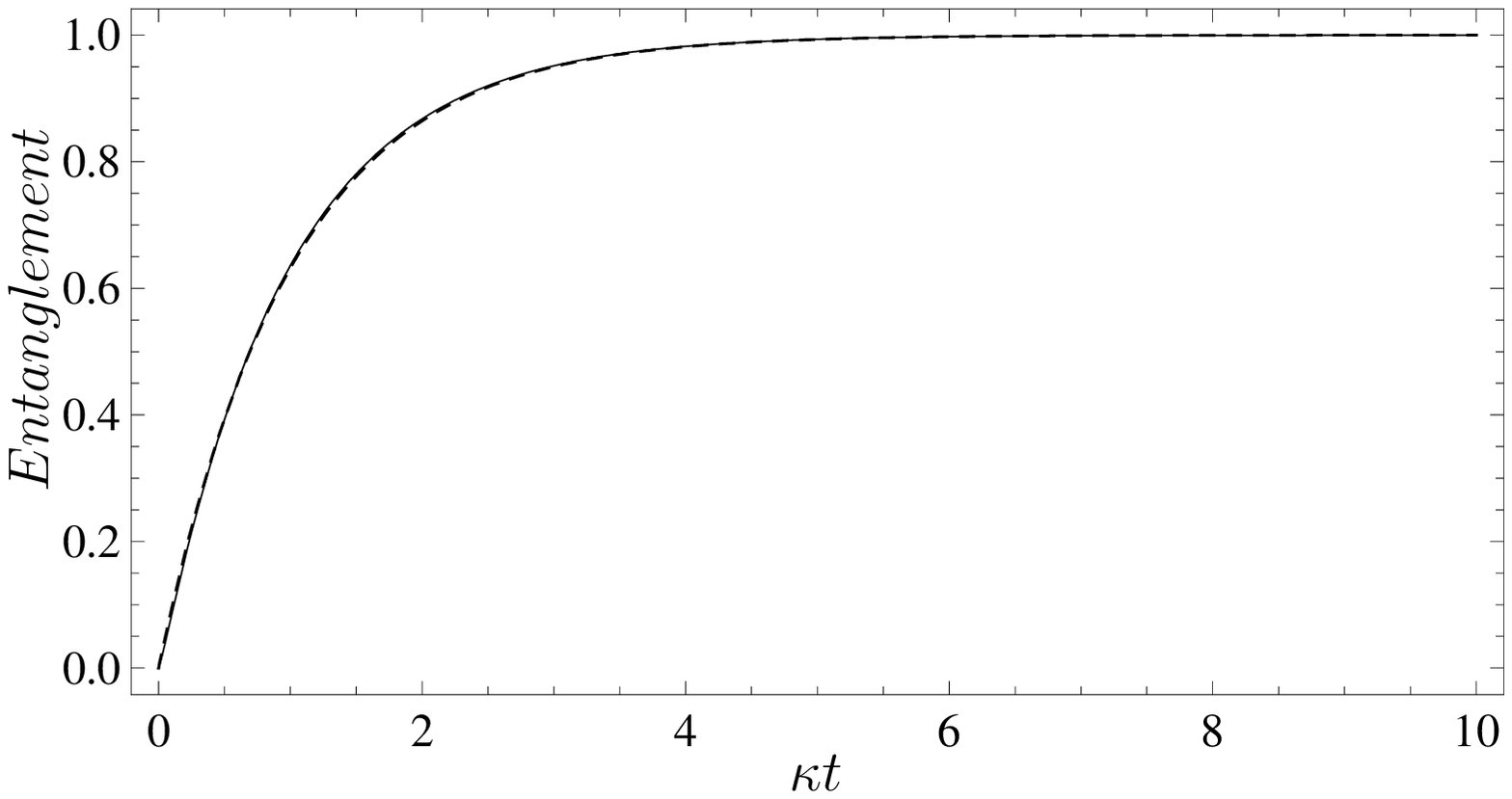}
\includegraphics[width=0.8\columnwidth]{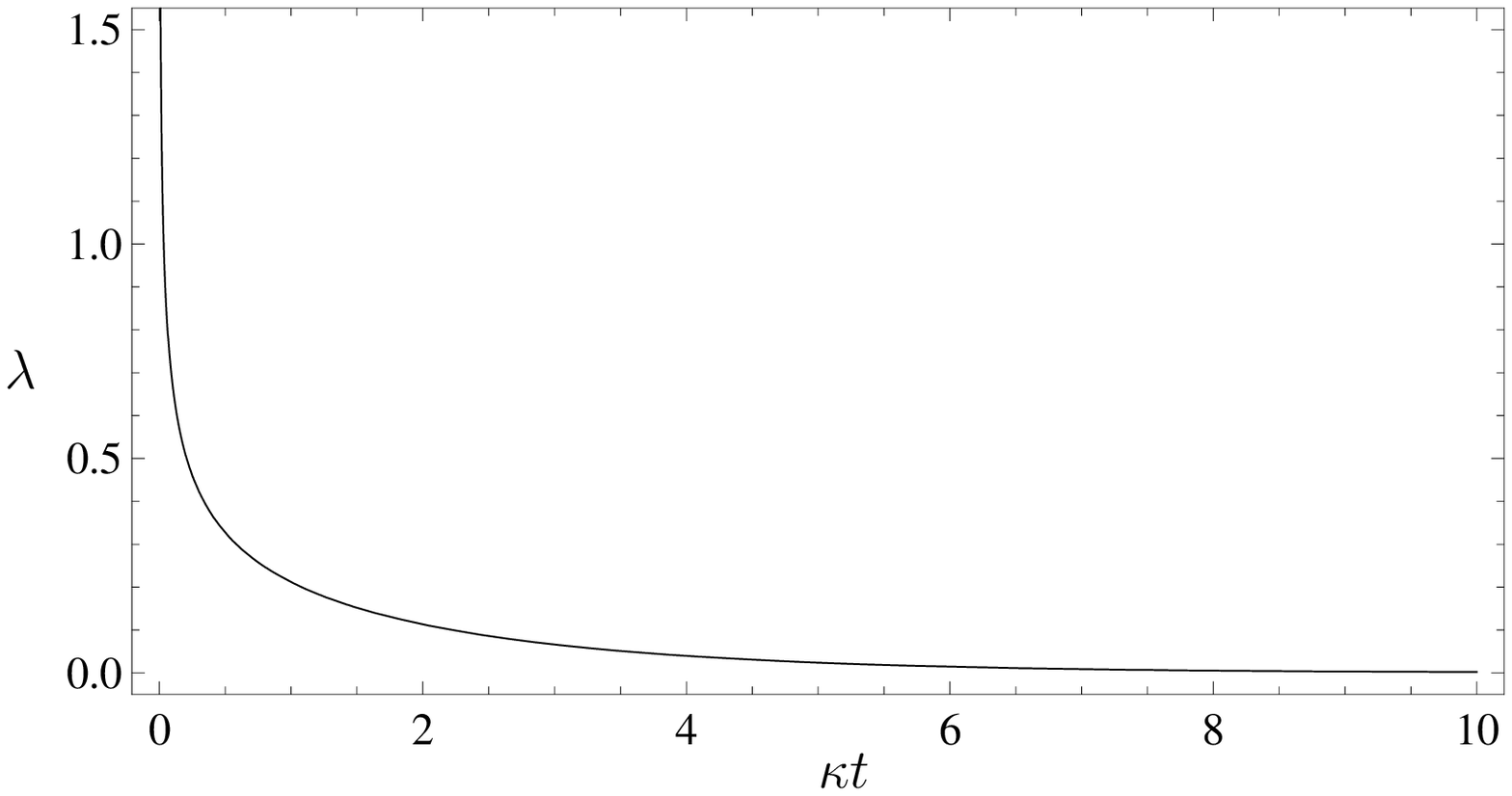}\\
\end{center}
\caption{\label{fig2} Example of temporal control of entanglement. (upper) The target function is plotted using dotted lines, and the designed entropy of entanglement is plotted using solid lines. (bottom) The time evolution of the coupling leading to the designed entanglement.}
\end{figure}

In Fig.\ref{fig2}, we show an example of entanglement design where we choose the entanglement temporal shape to be $f(t)=1-e^{-\kappa t}$. One can see that there is a very good agreement between $S(t)$ (the designed entanglement) and $f(t)$ (the target entanglement). This is in accordance with the successful linearization shown in Fig.\ref{fig1}. However, this example brings out an important fact, namely the coupling constant diverges at times when the entanglement goes to zero (in this example, at $t=0$). In fact, we can see that $\lambda(t)$ given by (\ref{c}) will diverge whenever $f(t)=0$ (because $q/2<1$) or $f(t)=1$. These divergences seem to be consequences of approximations. It is unlikely that the exact linearization (if possible to find) would lead to such divergences of the coupling for well-behaved functions. Besides, the form of the entropy of entanglement and the unitarity of time evolution (causing $\eta(t)$ to be continuous) forbid the choice of discontinuous $f(t)$. In general,  physical instantaneous abrupt changes in entanglement result from measurements and not from unitary evolution.
In spite of these divergences, the ansatz has been proved to offer a very accurate solution of the problem. In the next subsection, we present a renormalized version of the ansatz which is able to attain similar levels of precision but with the advantage of leading to physical (finite) couplings. 
\subsection{Renormalization Algorithm}
\begin{figure}[h]
\begin{center}
\includegraphics[width=0.8\columnwidth]{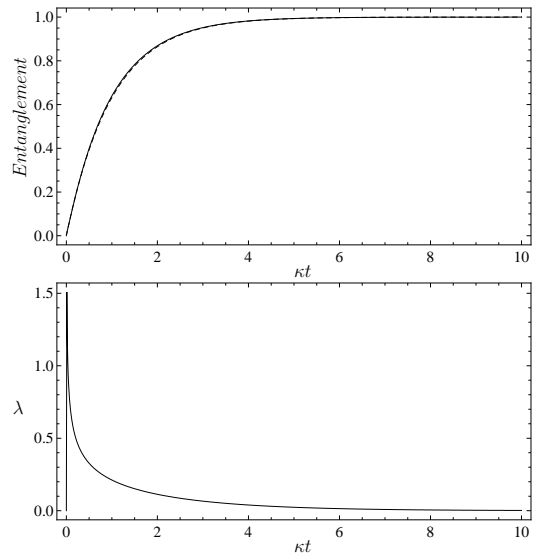}
\includegraphics[width=0.8\columnwidth]{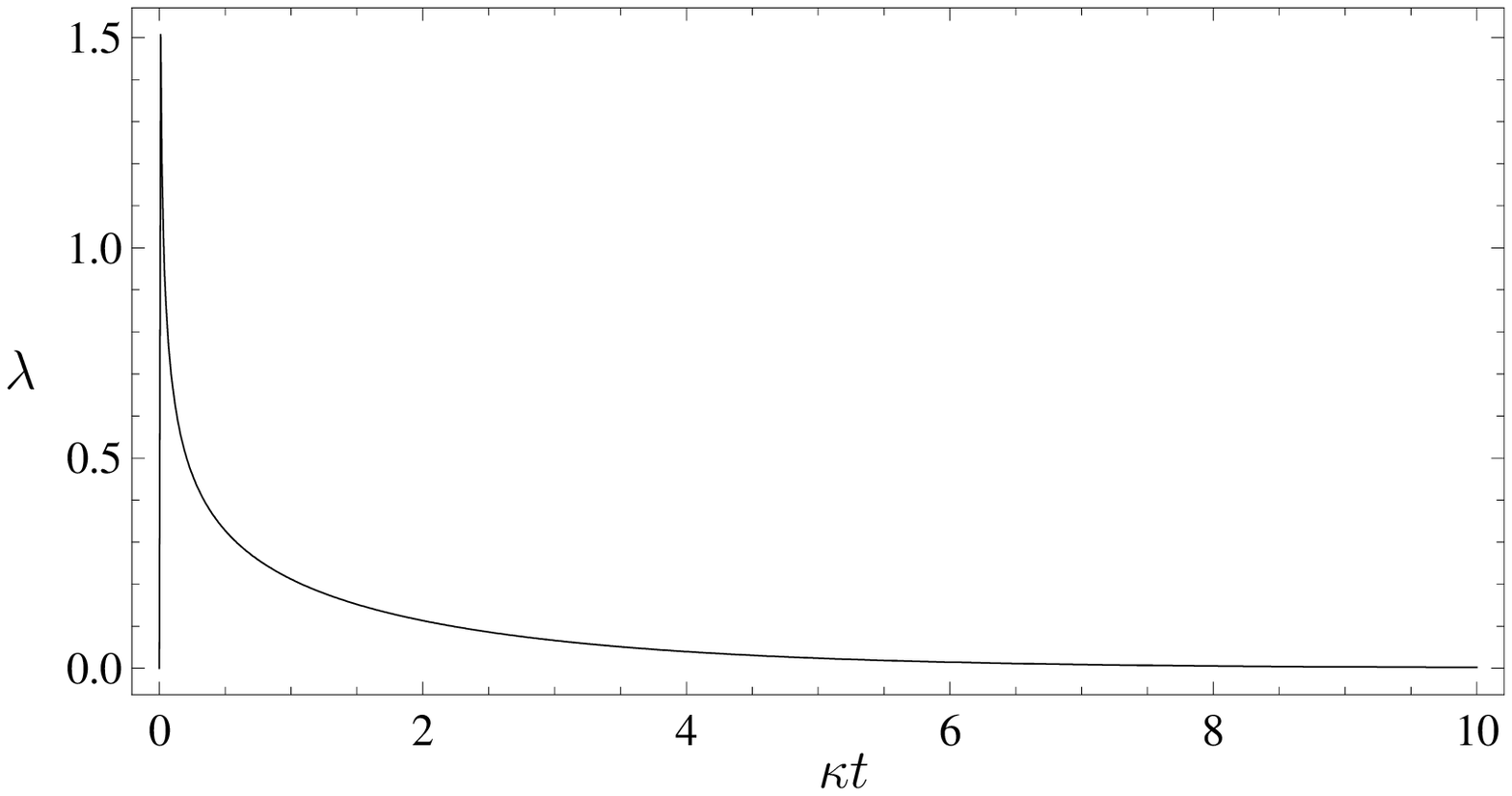}\\
\end{center}
\caption{\label{fig3} Example of temporal control of entanglement obtained using the renormalized ansatz. (upper) The target function is plotted using dotted lines, and the designed entropy of entanglement is plotted using solid lines. (bottom) The time evolution of the coupling leading to the designed entanglement.}
\end{figure}

 As a first attempt to eliminate the divergences in $\lambda(t)$, one could think of considering $q\geq 2$. This would certainly remove the divergence caused by $f(t)^{-1+q/2}$ around $f(t)=0$, but this would not help to solve the difficulties that arose at $f(t)=1$. Besides, as seen from the distance function plotted in Fig.\ref{figq}, the precision of the solution would be poor. A more effective idea is to keep $q<2$, actually keep the optimum value $q=1.345$, and then cut off the divergence of $\lambda(t)$ whenever it appears, i.e., around $f(t)=0$ and $f(t)=1$. This lead to the simple renormalization algorithm
\begin{eqnarray}
\lambda(t)&=&  \frac{1}{4}\frac{qf(t)^{-1+q/2}}{\sqrt{1-f(t)^q}}\frac{df(t)}{dt}\,\,\,\,\,\,{\rm{if}}\,\,\,\,\,\,\delta_0\leq f(t)\leq\delta_1\nonumber\\
\lambda(t)&=&\lambda_0 \,\,\,\, {\rm{otherwise}},\label{ran}
\end{eqnarray}
with $\lambda_0$ finite, $\delta_0>0$, $\delta_1<1$, and $q=1.345$. Fig.\ref{fig3} shows the renormalized version of the previous example, with the choices $\lambda_0=0$, $\delta_0=10^{-3}$, and $\delta_1=1-\delta_0$. The parameters $\lambda_0$, $\delta_0$, and $\delta_1$ can be chosen as to minimize the error while keeping $\lambda(t)$ finite, but we leave optimization to be studied elsewhere. As one can see from Fig.\ref{fig3}, there is still an excellent agreement between the target curve and the one obtained with our renormalized ansatz (\ref{ran}).

\begin{figure}[h]
\begin{center}
\includegraphics[width=0.8\columnwidth]{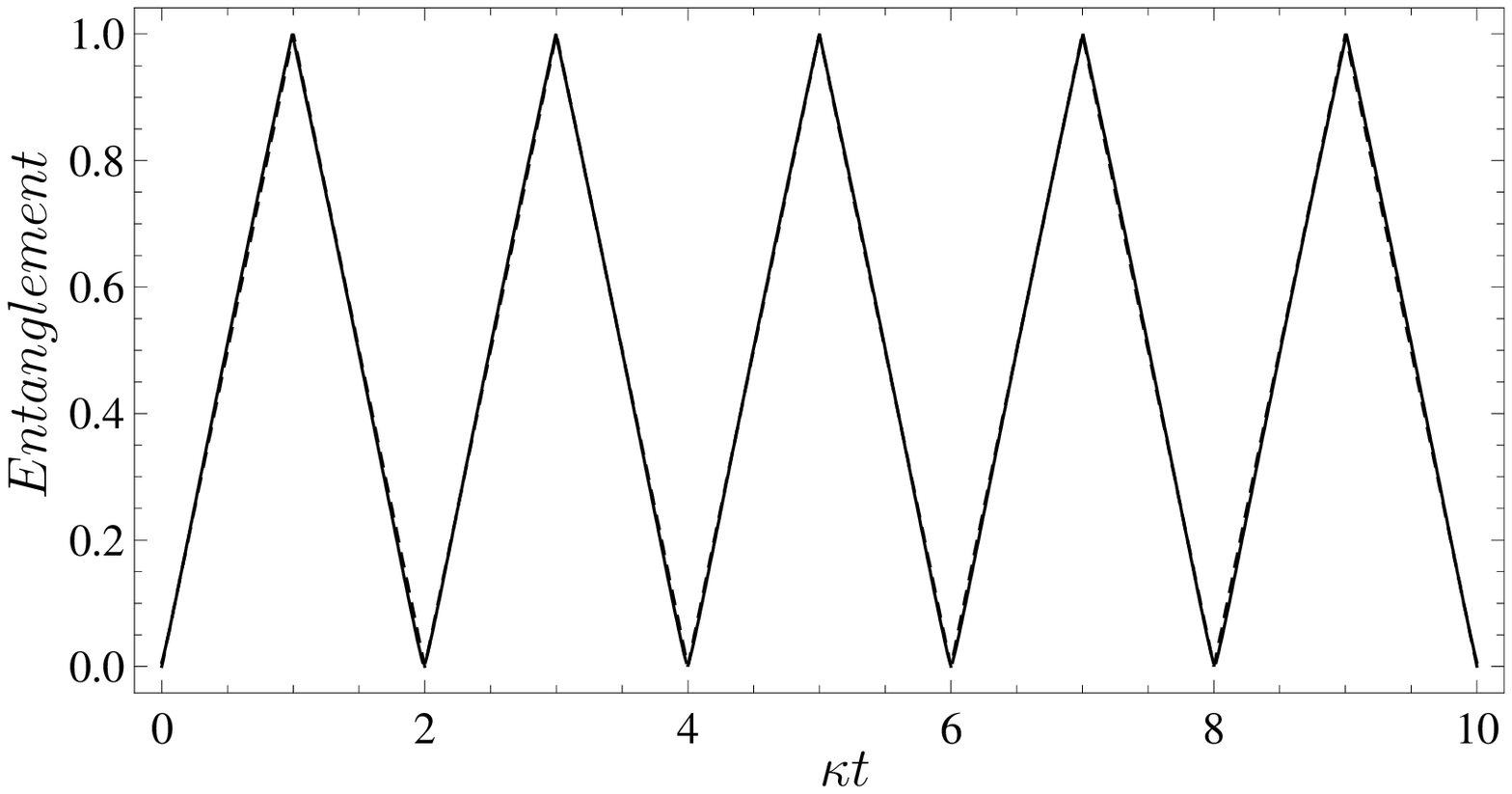}
\includegraphics[width=0.8\columnwidth]{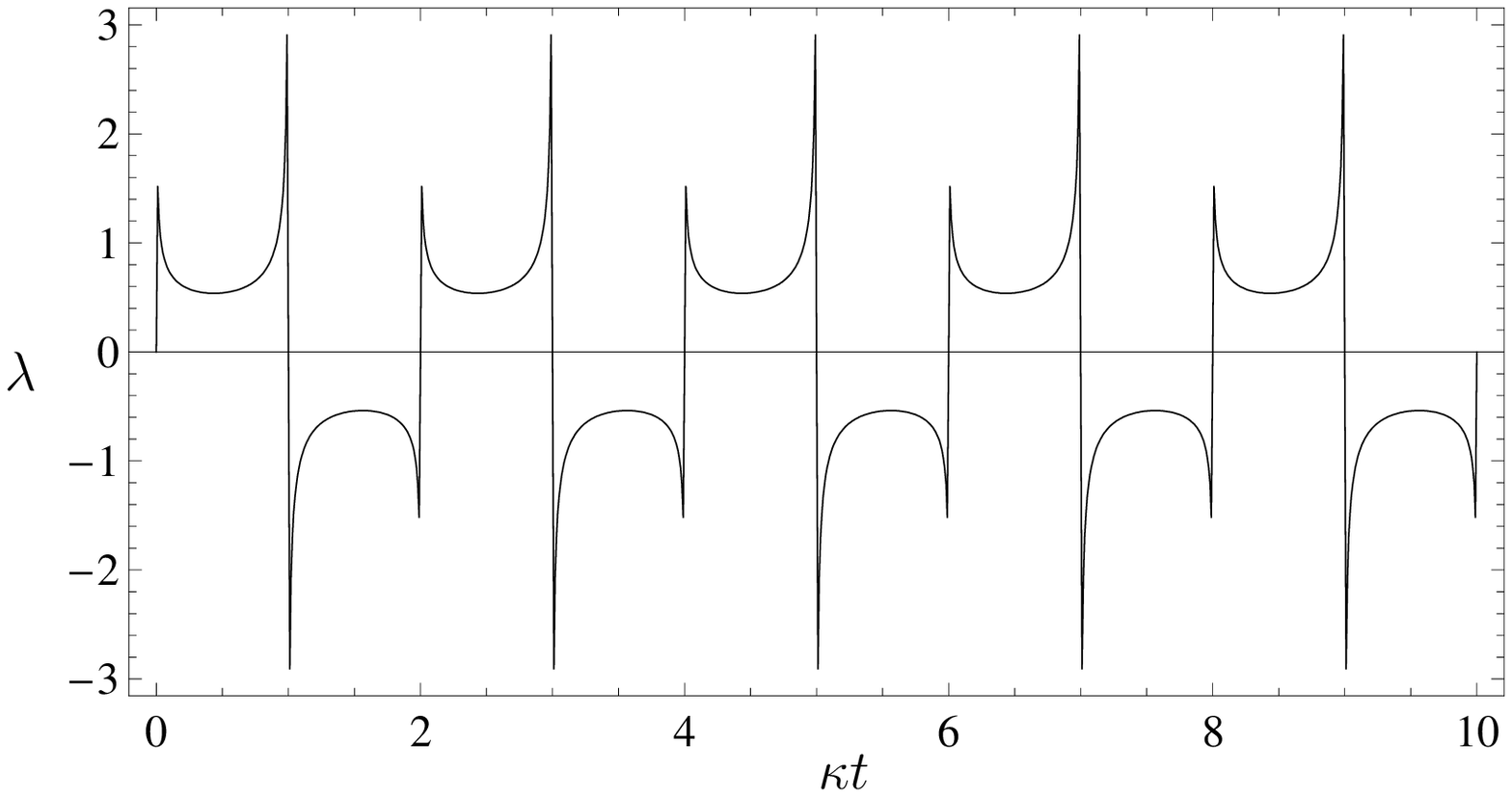}
\end{center}
\caption{\label{fig4}Example of temporal control of entanglement obtained using the renormalized ansatz.  (upper) The target function is plotted using dotted lines, and the designed entropy of entanglement is plotted using solid lines. (bottom) The time evolution of the coupling leading to the designed entanglement.}
\end{figure}

In Fig.\ref{fig4}, we provide another example of active temporal control of the entanglement evolution. The target function is now the \emph{exotic} $f(t)=\tfrac{1}{2}+\tfrac{1}{\pi}\arcsin[\sin(\pi\kappa  t-\pi/2)]$. Again, there is a very good precision in the design of the entanglement. In both plots, we used $q=1.345$, $\lambda_0=0$, $\delta_0=10^{-3}$, and $\delta_1=1-\delta_0$.

Our intention in this paper is not to treat optimization problems like minimization of the time taken for the system to achieve a certain amount of entanglement. Instead, we show that the system may have its entanglement following a chosen path between two fixed values. The Hamiltonian considered in this work as well as initial system preparation fix the starting point to be $0$ \emph{ebit} and allow the end point to be \emph{any value}, including $1$ \emph{ebit}. 
\subsection{Possible physical setting}
We would like now to analyze the feasibility of our idea in experimentally accessible physical systems. Experimental and theoretical work on superconducting qubits for quantum information purposes is a very fast growing field of investigation \cite{squbitsf,squbits,njp_qubits}. It is currently possible, for example, to experimentally couple two superconducting flux qubits through an Ising-type Hamiltonian \cite{science01,science02,exp1}
\begin{eqnarray}
H=-\sum_{i=1}^2\frac{1}{2}[\epsilon_i\sigma_i^z+\Delta_i\sigma_i^x]+J\sigma_1^z\sigma_2^z,\label{Is}
\end{eqnarray}
where $\epsilon_i$ and $\Delta_i$ represent the bias and tunneling energies of qubit $i$, respectively, and $J$ is the qubit-qubit coupling constant. 

There have been reported many different couplers between flux qubits providing a good control of the sign and magnitude of the $J$ coupling \cite{science01,exp1,exp2,exp3}.  In \cite{njp_qubits}, for example, it is theoretically shown how a monostable rf- or dc-SQUID can mediate coupling between two adjacent flux qubits described by Hamiltonian (\ref{Is}), and the experimental implementation of this scheme is reported in \cite{exp1}. In this experiment, measurements demonstrate that $J$ can be sign and magnitude tuned with a properly chosen flux bias in the monostable rf SQUID. This tunable coupling, added to the fact that flux waveforms can be electronically controlled, may be used  to obtain an arbitrary time-dependent coupling $J(t)$, the key ingredient needed in our scheme. In particular, the experiment reported in \cite{exp3} presents an improved tunable coupling element (compound Josephson junction rf-SQUID) able to provide a sign and magnitude tunable $J(t)$ with minimal nonlinear crosstalk from the coupler tuning parameter into the qubits.

For our purposes, Hamiltonian \ref{Is} is as good as (\ref{H}) as long as the tunelling energy is taken equal zero, and the qubits are initially prepared in the state $|\psi(0)\rangle=|{+}{-}\rangle$, where $|\pm\rangle$ are eigenstates of the Pauli matrix $\sigma_x$. In this case, the interaction picture evolved state considering Hamiltonian (\ref{Is}) will be 
\begin{eqnarray}
|\psi(t)\rangle=\cos\eta(t)|{+}{-}\rangle-i\sin\eta(t)|{-}{+}\rangle,\label{evx}
\end{eqnarray}
which is locally equivalent to (\ref{ev}). Consequently, the qubit-qubit entanglement is still given by (\ref{ee}) and our scheme can then be applied with no modifications. The case $\Delta_{1,2}\neq 0$ may work too by including application of microwave drivings \cite{science02}.
\section{Interpretation of the Entanglement Paths}\label{decoherence}
The question that naturally arises is: \emph{why would one want to obtain control over entanglement trajectories?} This section is dedicated to a possible answer to this question.  Let us consider the region defined by the ordained pairs $\{(\kappa t,Entanglement)\in [0,10]\times[0,1]\}$, just like considered in the plots shown in Figs.(\ref{fig3}) and (\ref{fig4}). The curve $R(t)=(\kappa{}t/10)$ divides the plane into two different sets according to entanglement: trajectories above this curve mean that the system visits states with higher entanglement, while going from $0$ to $1$ \emph{ebit}, than when $f(t)<R(t)$. In the ideal case, it is always possible to achieve $1$ \emph{ebit}, independently on the trajectory taking place above or below the boundary defined by $R(t)$. As we are going to show, different trajectories controlled by using our ansatz provide a direct visualization of the effect of the decoherence acting on the qubits. This gives us a first clue about the meaning and application of different choices of $\lambda(t)$ (\ref{ran}) for particular user-defined entanglement trajectories $f(t)$ originally defined in the ideal case. 
   
In the previous section, we presented the qubit-qubit dynamics following a designed coherent evolution to achieve a final state with  maximum amount of entanglement. We now analyze the problem in the presence of two representative types of decoherence channels: the amplitude-damping (AD) and  phase-damping (PD) channels. The description of the incoherent dynamics of a composed system is usually made by using the appropriate master equations in the Lindblad form. The derivation of these equations are discussed with great detail in \cite{carmichael}. The density matrix $\rho(t)$ describing the system state under Born-Markov approximations obeys the master equation in the Lindblad form
\begin{eqnarray}\label{me}
\dot\rho(t)&=&-i[H(t),\rho(t)]+\sum_k[L_k\rho(t)L^\dag_k-\frac{1}{2}L^\dag_kL_k\rho(t)\nonumber\\ &&-\frac{1}{2}\rho(t)L^\dag_kL_k],
\end{eqnarray}
where the first term gives the unitary part of the time evolution while the others describes the irreversible coherence loss into the environment. The operators $L_k$ are the called Lindblad or quantum jump operators \cite{preskill}.

The AD channel describes the dissipative dynamics where the qubit loses excitation to the environment, at decay rate $\Gamma$. Thus, each qubit dissipative process is described by a single Lindblad operator $L_k=\sqrt{2\Gamma}\sigma^-_k$, with $\sigma^-_k=(\sigma^x_k-i\sigma^y_k)/2$ ($k=1,2$). Substituting these jump operators in (\ref{me}), we have the system master equation under the AD channel
\begin{eqnarray}\label{meAD}
\dot\rho(t)&=&-i[H(t),\rho(t)]+\sum_{k=1,2}\Gamma[2\sigma^-_k\rho(t)\sigma^+_k-\sigma^+_k\sigma^-_k\rho(t)\nonumber\\ &&-\rho(t)\sigma^+_k\sigma^-_k].
\end{eqnarray}

The PD channel is a purely quantum channel leading to losses of quantum coherence without any energy relaxation. In this channel, each qubit suffers the action of a single Lindblad operator $L_k=\sqrt{\Gamma}\sigma^z_k$ ($k=1,2$). By substituting these jump operators in (\ref{me}), we have the master equation for the PD channel
\begin{eqnarray}\label{mePD}
\dot\rho(t)=-i[H(t),\rho(t)]+\sum_{k=1,2}\Gamma[\sigma^z_k\rho(t)\sigma^z_k-\rho(t)].
\end{eqnarray}

To quantify the entanglement content of mixed states, we consider the entanglement of formation, which is defined as \cite{eof}
\begin{eqnarray}\label{eof}
EoF=h[(1+\sqrt{1-C^2})/2),
\end{eqnarray}
where $h(x)=-x\log_2x-(1-x)\log_2(1-x)$ and $C$ is the concurrence. For the initial preparation $|\psi(0)\rangle=|01\rangle$, the density operator obeying the master equations (\ref{meAD}) or (\ref{mePD}) fall into the class of what is called $X$ states \cite{Xstates}. For these very symmetrical states,  concurrence assumes the simple form

\begin{eqnarray}
C&=&2\max\{0,|\rho_{23}(t)|-\sqrt{\rho_{11}(t)\rho_{44}(t)},|\rho_{14}(t)|\nonumber\\ &&-\sqrt{\rho_{22}(t)\rho_{33}(t)}\},
\end{eqnarray}
where $\rho_{ij}(t)$, with ($i,j=1,\ldots,4$), are the corresponding matrix elements of $\rho(t)$ in the computational basis.

\begin{figure}[h]
\includegraphics[width=0.8\columnwidth]{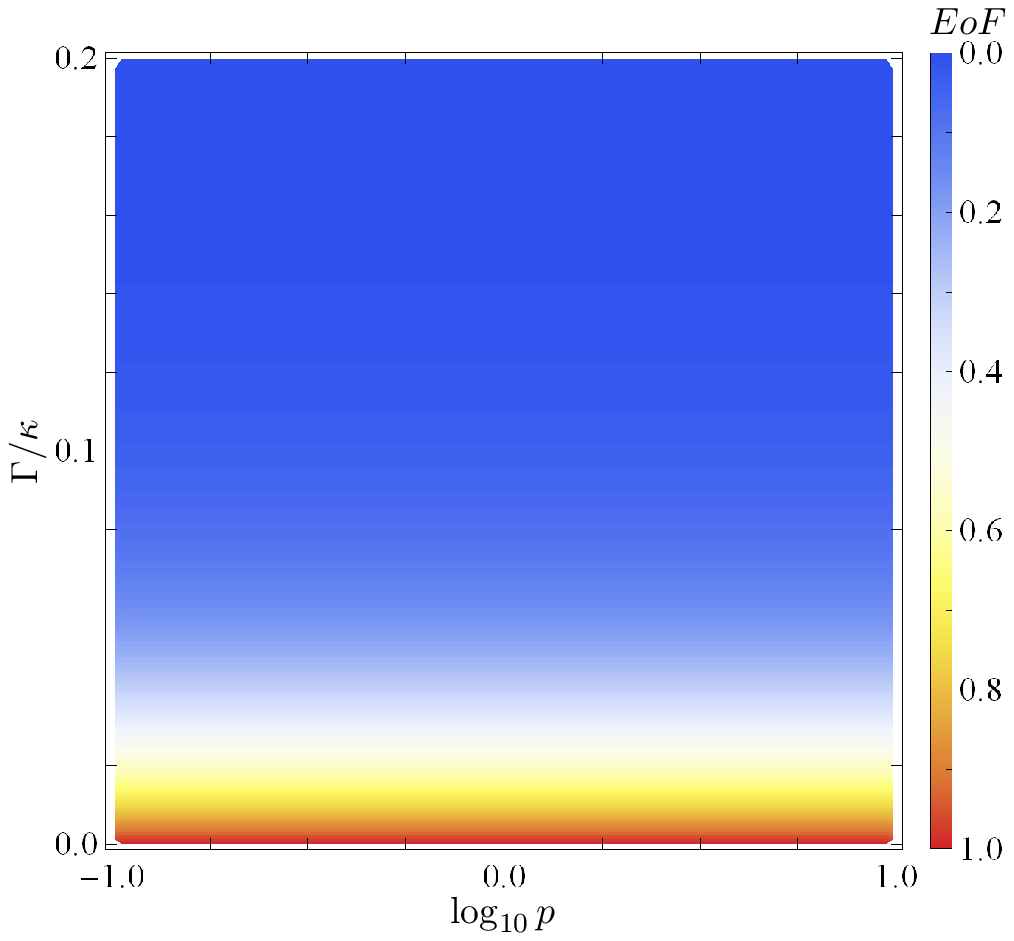}
\includegraphics[width=0.8\columnwidth]{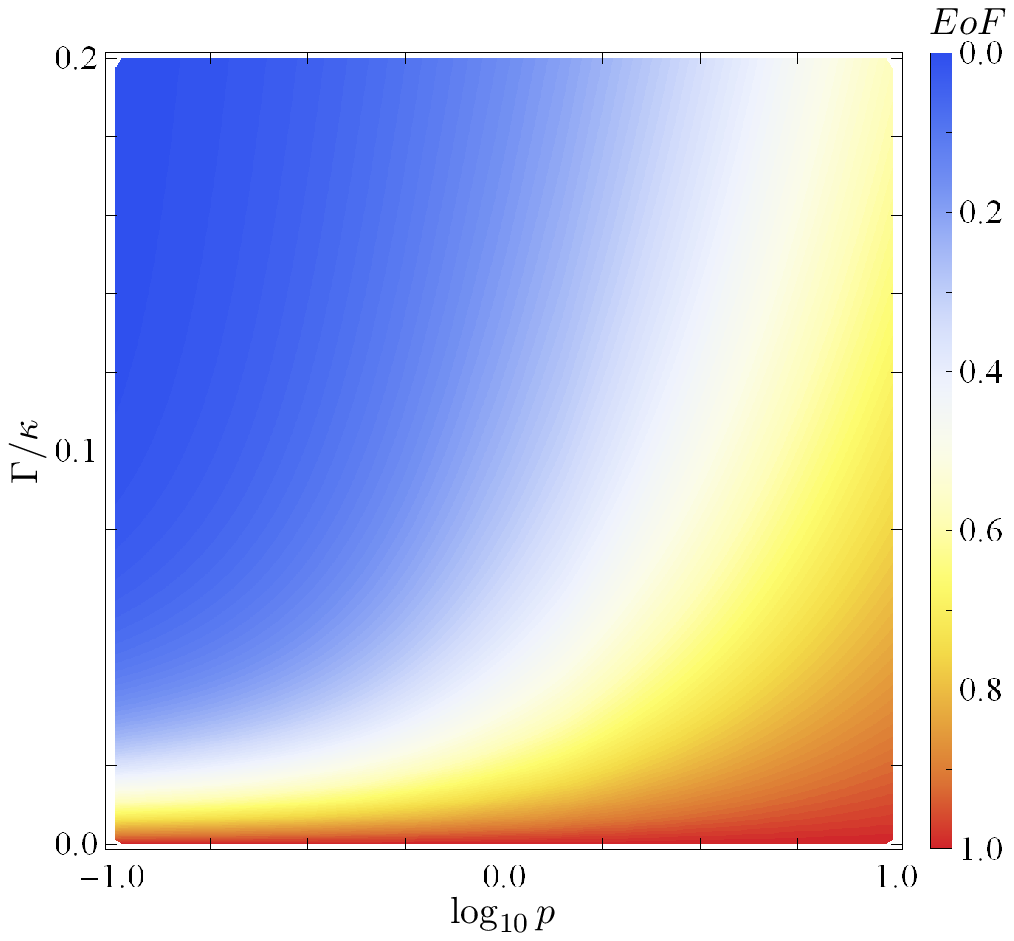}
\caption{Final amount of entanglement achieved when following a path given by $f(t)=(\kappa{}t/10)^p$ in the presence of the AD channel (top)  and PD channel (bottom), as a function of $\Gamma/\kappa$ (vertical axis) and ($\log_{10}p$) (horizontal axis).}
\label{fig:alpha}
\end{figure}

We are know in position to get back to the original question. In order to answer it, we will consider what happens with final amount of entanglement achieved under the influence of these decoherence channels, when different prescribed couplings $\lambda(t)$, originally obtained from user-defined functions $f(t)$ ranging from $0$ to $1$ \emph{ebit}, are calculated with our ansatz (\ref{ran}).
We define a family of functions $f(t;p)=(\kappa{}t/10)^p$, where the parameter $p$ selects the region delimited by $R(t)=f(t;1)$. If $p>1$ ($p<1$) the path in the ideal case would access states less (more) entangled before reaching $1$ \emph{ebit}, as delimited by $R(t)$. In Fig. \ref{fig:alpha}, we show the final amount of entanglement achieved when following a path given by $f(t;p)$ under the influence of each decoherence channel. From these plots one can see that the AD channel does not remove the \emph{degeneracy} of symmetric paths $p$ and $1/p$. The situation changes radically for the PD channel, where one can clearly see that it is more advantageous to spend most of the time in states with less entanglement.
\section{Final Remarks}\label{conclusions}
To summarize, we have proposed a method to control the time evolution of entanglement between two qubits in real time. In order to achieve such control, we proposed the use of time-dependent coupling magnitudes, and provide a relation between the interaction area pulse or coupling constant wave form and the desired shape of the entanglement function. These entanglement trajectories, under specific conditions, revealed an independence of the final entanglement on each of the symmetric trajectories (defined in the text) for the case of amplitude damping. The phase damping channel has been shown to not possess this symmetry. Consequently, our scheme for controlling $\lambda(t)$ may find applications in experiments aiming at entanglement maximization using XY Hamiltonians in the presence of the PD channel.  We hope this work can trigger studies including generalizations for more involved systems such as general spin chains or coupled continuous variable systems. 

\begin{acknowledgments}
F.L.S. thanks F. Brito for discussions related to this work. We acknowledge partial support from the Brazilian National Institute of Science and Technology of Quantum Information (INCT/IQ). P.P.M. thanks CAPES/CNPq under grant PNPD0030082. F.L.S. also acknowledges partial support from CNPq under grant 303042/2008-7.
\end{acknowledgments}


\begin{thebibliography}{xxxxx}
\bibitem{squbitsf} Our choice for one or other theoretical scheme or experimental implementation does not reflect any bias of choice or preference for any particular research group. Our intention is to illustrate the feasibility of our proposal using current technology.
\end{thebibliography}
\end{document}